\documentclass[a4paper]{article}

\usepackage{INTERSPEECH2021}
\usepackage{cite}
\usepackage{amsmath,amssymb,amsfonts}
\usepackage{algorithmic}
\usepackage{graphicx}
\usepackage{textcomp}
\usepackage{xcolor}
\usepackage{multirow}
\usepackage{arydshln}
\usepackage{color}

\title{An evaluation of data augmentation methods for sound scene geotagging}
\name{Helen L. Bear$^1$, Veronica Morfi$^1$, Emmanouil Benetos$^{1,2}$}
\address{
  $^1$Centre for Digital Music, Queen Mary University of London, London, UK\\
  $^2$The Alan Turing Institute, London, UK}
\email{\{h.bear, g.v.morfi, emmanouil.benetos\}@qmul.ac.uk}

\begin{document}

\maketitle
\begin{abstract}
Sound scene geotagging is a new topic of research which has evolved from acoustic scene classification. It is motivated by the idea of audio surveillance. Not content with only describing a scene in a recording, a machine which can locate where the recording was captured would be of use to many. In this paper we explore a series of common audio data augmentation methods to evaluate which best improves the accuracy of audio geotagging classifiers. 

Our work improves on the state-of-the-art city geotagging method by $23\%$ in terms of classification accuracy.
\end{abstract}


\section{Introduction}
Sound scene geotagging is a relatively novel topic of research. It means to correctly identify the geographical position where a recording was captured \cite{bear2019city} and as such it can also be described as either audio geotagging or as a more specific task such as audio-based city classification. Subject to the precision of the dataset annotations, it could be formulated as a coarse target problem (e.g. at the level of a city) or an investigation of an even more precise target (e.g. at the level of coordinates or postcodes), if such labels are provided. To the best of our knowledge no such dataset exists with labels on a more granular level than cities. Hence, for this paper, city classification from sound scenes is the primary geotagging task. 

Sound scene geotagging is a task with many future real-world applications, such as automatically knowing a precise geo-location from the background sounds of an emergency call that could potentially decrease the response time~\cite{1416352} or it could position a person in a location with no cameras at a specific time for criminal investigations. However, the current state-of-the-art in audio geotagging that uses multi-task learning, achieves only $56\%$ accuracy \cite{bear2019city} on city classification.

In this paper, we seek to improve the state-of-the-art performance in sound scene geotagging. One approach is to train on more data. However, data collection is expensive, and more so when one would have to travel to collect recordings from different cities. Thus, we seek to achieve a dataset expansion by using a series of data augmentation methods. 

Data augmentation is a common way of improving accuracy scores in deep learning methods. In audio some examples are: genre classification~\cite{aguiar2018exploring}, speech recognition~\cite{ko2015audio}, and sound event detection~\cite{chen2019rare}. In this work, we explore a series of common audio data augmentation methods to evaluate which ones best improve city classification accuracy. This work is conducted using data which is varied in the types of scenes recorded in each possible city. Through our experiments we found that augmentations that can preserve all the information of the original signal, such as time shift (referred to as \textit{cyclic} augmentation in this manuscript), provide the highest improvement on the performance of our sound scene geotagging method. 

The rest of this paper is as follows: a short background in Section \ref{sec:background} discussing prior work is followed by Section \ref{sec:data} presenting the data available for sound scene geotagging and the augmentation methods used in our analysis. Section \ref{sec:method} follows describing our methodology and the results. Finally, Section \ref{sec:conclusions} concludes this paper with our observations on these experiments and possible future work.

\section{Background}
\label{sec:background}
Prior to the work in \cite{bear2019city}, there was little investigation of audio geotagging, and to the best of our knowledge, not as a classification task. Examples by Elizalde et al. \cite{7545041} and Kumar et al. \cite{kumar2017audio} used sound scenes generated with sound events not specific to the locations. This seems non-intuitive for city classification, as the sound events which are unique to a city would be easily learnt by a classifier. In addition to this, in \cite{7545041} and \cite{kumar2017audio} the problem of geotagging is not formulated as a classification task but rather as retrieval task seeking the most similar city in a clustering model.

Whilst sound scene geotagging has received limited attention, there is work which uses audio to help geotag videos for visual scene understanding or in multi-media systems \cite{luo2011geotagging}. In videos it can be rather straight-forward to identify visual features in a scene that correspond to audio tracks and use those for geotagging a location. Estimated geo-tags can be used as metadata which enables better prediction of events in videos \cite{joshi2008inferring}. Furthermore, they can be used for modelling relationships between spatio-temporal segments and entities in multimedia data \cite{luo2006pictures}.

In \cite{bear2019city}, the authors use a dataset labelled for both audio geotagging (in a city level) and acoustic scene classification. As such, the annotations function as a many-to-many connection between the city and scene classes. Each recording maps to both one scene and one city. This results in multiple types of scenes appearing in a single city and multiple cities containing each scene. Both labelling schemes are based on how humans describe a scene or identify a location~\cite{bear2018extensible}. The fact that the labels are not data driven means there can be features in the recordings which are confounding variables that a classifier has to learn to discriminate~\cite{heise2020fuzz}. This dataset is also used in our experiments and it is described in the following section along with the augmentation methods applied on the recordings.

\begin{table*}[!h]
    \centering
    \caption{The CNN structure and parameters for a bigger model able to cope with the greater volume of training data. Adam optimiser with $LR=0.0001$, $beta_1=0.9$, $beta_2=0.999$, $epsilon=None$, $decay=0.0$, $amsgrad=False$.  }
    \begin{tabular}{r r r |l l l}
    \hline
     & & Layer & params \\
    \hline \hline
    1 & \multicolumn{2}{r}{Convolutional2D}     & \multicolumn{2}{l}{filters=64, kernel=(7,7)}  \\
    2 & \multicolumn{2}{r}{BatchNormalization}  & \multicolumn{2}{l}{}\\
    3 & \multicolumn{2}{r}{MaxPooling2D}        & \multicolumn{2}{l}{pool\_size=(5,5), strides=2, padding=`same'} \\
    4 & \multicolumn{2}{r}{Dropout}             & \multicolumn{2}{l}{prob\_drop\_conv=0.3} \\
    
    5 & \multicolumn{2}{r}{Convolutional2D}     & \multicolumn{2}{l}{filters=128,  kernel=(7,7)}   \\
    6 & \multicolumn{2}{r}{BatchNormalization}  & \multicolumn{2}{l}{}\\
    7 & \multicolumn{2}{r}{MaxPooling2D}        & \multicolumn{2}{l}{pool\_size=(4,4), strides=2, padding=`same'} \\
    8 & \multicolumn{2}{r}{Dropout}             & \multicolumn{2}{l}{prob\_drop\_conv=0.3} \\
    
    9 & \multicolumn{2}{r}{Convolutional2D}     & \multicolumn{2}{l}{filters=256, kernel=(5,5)}   \\
    10 & \multicolumn{2}{r}{BatchNormalization} & \multicolumn{2}{l}{}\\
    11 & \multicolumn{2}{r}{MaxPooling2D}       & \multicolumn{2}{l}{pool\_size=(3,3), strides=2, padding=`same'} \\
    12 & \multicolumn{2}{r}{Dropout}            & \multicolumn{2}{l}{prob\_drop\_conv=0.3} \\
    
    13 & \multicolumn{2}{r}{Convolutional2D}    & \multicolumn{2}{l}{filters=512, kernel=(3,3)}   \\
    14 & \multicolumn{2}{r}{BatchNormalization} & \multicolumn{2}{l}{}\\
    15 & \multicolumn{2}{r}{MaxPooling2D}       & \multicolumn{2}{l}{pool\_size=(2,2), strides=2, padding=`same'} \\
    16 & \multicolumn{2}{r}{Dropout}            & \multicolumn{2}{l}{prob\_drop\_conv=0.3} \\
    \hdashline
    \multicolumn{5}{c}{layer 16 output feeds into two separate output blocks, 17 and 19}. \\
    \hdashline
    17 & \multicolumn{1}{r}{GlobalAveragePooling2D}& \\
    18 & \multirow{2}{*}{Dense} & filters=10, activation=`sigmoid', \\
        &                    & kernal\_regularizer = L2($0.001$) \\
 
    19 & &  & GlobalAveragePooling2D & \\
    20 & &  & \multirow{2}{*}{Dense} & filters=6, activation=`sigmoid', \\
    & &     &                    & kernal\_regularizer = L2($0.001$) \\
    \hline
    \end{tabular}   
    \label{tab:bigparams}
\end{table*}

\section{Data}
\label{sec:data}
The DCASE 2018 Acoustic Scene Classification (ASC) subtask 1A \cite{Mesaros2018_DCASE} contains recordings from six cities: Barcelona, Helsinki, London, Paris, Stockholm, and Vienna. The training data partition consists of approximately 70\% of the recordings from each city. Recordings include ten predefined scenes~\cite{Mesaros2018_DCASE} at random times between 9am and 9pm on different weekdays. The development set consists of 8640 audio segments in total, split into a training and an evaluation subset containing 6122 and 2518 segments, respectively. In our experiments, we use the evaluation subset for validation. The test set consists of 2518 audio segments, previously held back in order to review the performance of the DCASE challenge submissions. The development dataset contains in total $24$ hours of audio, while the testing set contains in total $7$ hours of audio. The equipment used for recording consists of a binaural Soundman OKM II Klassik/studio A3 electret in-ear microphone and a Zoom F8 audio recorder using $48$kHz sampling rate and $24$-bit resolution.

\subsection{Data Preparation}
\label{sec:augments}
Before any data augmentation is performed, we first downsample the audio files to 22050Hz. Then, in order to enhance our training set, we apply three different data augmentation methods, two on the waveforms and one on the time-frequency representations (of FFT window length of 2048), followed by trimming the lowest 100Hz and highest 100Hz frequencies. Finally, we compute the log mel-spectrogram (128 mel bands), which will be used as input to our network. Augmentations are applied only on the training set, while the validation and testing sets are kept the same. 
 
\begin{enumerate}
    \item Waveform augmentations: \\
         \textbf{Cyclic}. We shift each audio sample in time by 50\% of its length. In a waveform this means that we cut it into two parts, at 50\% of its length, and place the second part in front of the first. This augmentation is able to preserve all the information of the original waveform. Cyclic augmentation produces one new audio sample per input audio sample.\\ \\
        \textbf{Drop}. We perform time interval dropout by skipping a random number of samples (between 1 and length of recording) in a random index in the waveform. We do this twice per audio segment, using different random numbers of skipped samples and different random indices in the waveform. Drop augmentation produces two new audio samples per input audio sample.
    \item Spectrogram augmentations: \\
        \textbf{Stretch}. We apply time and frequency stretching by resizing a random number of columns and rows at a random position followed by image resizing using bilinear interpolation. This is performed four times on each audio segments with different random number of columns and rows to stretch and stretch factor. Stretch augmentation produces four new samples per input audio sample.
\end{enumerate}

\begin{table}[!th]
  \caption{Training data variation by data augmentation method.}
  \label{tab:trainData}
  \centering
  \begin{tabular}{l c  }
    \hline
    \textbf{Augmentation} & \textbf{\#Training Samples}  \\
    \hline
    none    & $6,122$    \\
    \hline
    cyclic  & $6,122$  \\ 
    drop    & $12,244$ \\
    stretch & $24,488$  \\
    \hline
    \textbf{Total}     & $48,976$  \\
    \hline
  \end{tabular}
\end{table}

The number of samples produced by each augmentation method can be found in Table \ref{tab:trainData}. These will be used in addition to the original $6122$ training samples for building CNNs.

\begin{table*}[!ht]
  \caption{City classification accuracy by data augmentation method with different CNN models.}
  \label{tab:results}
  \centering
  \begin{tabular}{l l | c c c c | c | r }
    \hline
    \multicolumn{7}{c}{Accuracy (\%) obtained with the multi-task  CNN from \cite{bear2019city} } \\
    \multicolumn{2}{c}{} &	cyclic	& drop & stretch  & 	cyclic+stretch	& all & benchmark  \\
    \hline
    \multirow{3}{*}{Scene} & train      & $85$ & $74$  & $86$ & $86$ & $83$ &  \\
                          & validation  & $46$ & $17$ & $49$ & $54$ & $55$ &   \\
                          & test        & $47$ & $16$ & $49$  & $53$ & $55$ &  $57$ \\
     \hdashline
     \multirow{3}{*}{City} & train      & $80$ & $64$& $82$  & $83$ & $82$ & \\
                          & validation  & $56$ & $30$ & $53$  & $55$ & $60$ &  \\
                          & test        & $56$ & $29$ & $52$  & $55 $ & $60$ &   $56$ \\
    \hline
    \multicolumn{7}{c}{Accuracy (\%) obtained with a larger multi-task CNN} \\
    \multicolumn{2}{c}{}&	cyclic	& drop & stretch  & cyclic+stretch	& all & benchmark  \\
    
    \hline
    \multirow{3}{*}{Scene} & train      & $99$ & $95$ & $99$  & $99$ & $99$ & \\
                          & validation  & $70$ & $19$ & $62$  & $67$ & $65$ & \\
                          & test        & $70$  & $19$ & $63$ & $67$ & $66$ & $57$ \\
    \hdashline
    \multirow{3}{*}{City} & train       & $99$ & $94$ & $99$ & $99$ & $99$ &  \\
                          & validation  & $79$ & $46$ & $75$ & $72$ & $73$ &  \\
                          & test        & \textcolor{red}{\textbf{79}} & $47$& $75$  & $71$ & $74$ & $56$\\
    
    \hline
    \multicolumn{7}{c}{Accuracy (\%) obtained with a larger single-task CNN} \\
    \multicolumn{2}{c}{}& cyclic &	drop	& stretch & cyclic+ctretch  & All & benchmark \\
    \hline
    \multirow{3}{*}{City}   & train      & $99$ & $99$ & $99$ & $99$ & $91$ &  \\
                            & validation & $76$ & $69$ & $72$ & $77$ & $59$ &  \\
                            &  test      & $75$ & $69$ & $71$ & $77$ & $59$ & $56$   \\
    \hline
  \end{tabular}
\end{table*} 

These waveform augmentations methods are widely used in audio applications when retaining as much information about the original signal as possible is important (such as for animal sounds \cite{NANNI2020101084}). Cyclic augmentation can enhance an audio dataset by preserving the original signal information, while drop creates more variety of new audio signals by removing time steps. On the other hand, stretch in the time-frequency domain is adjusted from the image stretching for augmentation. We assume that with small stretching factors the overall scene and city information can be preserved in the spectrogram.

\subsection{Mono vs Stereo}
During the process of applying the data augmentations we experienced some challenges worthy of note; our recordings are stereo two channel audio. We tested experimentally and learned that augmentations are more effective when they are applied to each channel separately (taking care to ensure that they are consistent for aspects such as temporal alignment) before the channels are averaged into a single channel feature matrix, rather than averaging before applying the augmentation.

\section{Experiments}
\label{sec:method}

We use three different CNN architectures. For the first one, we use the same multi-task model architecture as presented in \cite{bear2019city}. We train a different model for each augmentation method. Each model is trained on the original data and the selected augmented data; namely \textit{cyclic}, \textit{drop} and \textit{stretch}. The benchmark we seek to outperform is $56\%$ accuracy for city classification \cite{bear2019city}. Additionally, we train a model with all the augmentations that outperform the baseline and the original data (referred to as \textit{cyclic+stretch}). Finally, we train a model that includes all augmentations and the original data (referred to as \textit{all}). 

For the second architecture, the same series of experiments are repeated using a larger architecture than the baseline model. The parameters and architecture of this model are detailed in Table~\ref{tab:bigparams}. 
As the focus of our work is audio geotagging rather than acoustic scene classification, we also use a third architecture which does not perform multi-task predictions but solely focuses on the city classification task. The single-task architecture is similar to the multi-task model from Table ~\ref{tab:bigparams} with the scene classification branch being removed (layers 17 and 18). 

\section{Results}
\label{sec:results}
The results for all our experiments are reported in Table~\ref{tab:results}. Best performance on city classification is achieved with the larger CNN model with the use of only the cyclic augmentation. The results of the first architecture based on \cite{bear2019city} are shown in the top of the table. The results of the larger CNN architecture from Table~\ref{tab:bigparams}) are in the middle of the table and the final single-task CNN results are presented in the bottom of the table. 

With the architecture of \cite{bear2019city}, we struggle to significantly outperform the original benchmarks; for acoustic scene classification (ASC) none of the augmentations outperform the $57\%$ benchmark but for sound scene geotagging (city classification), we match the benchmark with cyclic augmentations and improve on it by $4\%$ by using all three. Using drop augmentations significantly reduces the performance, possibly due to the method removing the entirety of some shorter sound events that might be unique to particular locations. Further experiments using different drop factors could be implemented to test this hypothesis. We sought to outperform the all augments result by removing the drop augment in the cyclic+stretch trained model, however this was not successful by only scoring $55\%$, one percent less than the benchmark. 

With our larger CNN architecture, different behaviour is observed. Whilst drop augmentation is still detrimental compared to the benchmarks for both ASC and audio geotagging (city classification), both cyclic and stretch achieved large increases in accuracy, $+13\%$ and $+6\%$ respectively for ASC and $+23\%$ $+19\%$ for audio geotagging. The performance of cyclic augmentations for this model provides us with the best overall accuracy of $79\%$. Whilst the drop augmentation is still the lowest performing one, it is improved by the larger model architecture with $47\%$ test accuracy compared to the previous model with $29\%$ test accuracy. 

Finally, for the single-task CNN performing only city classification, all augmentation methods, including drop, achieve an accuracy greater than the benchmark. However, the model using all augmentation methods together has a lower performance than any of the models using single augmentations. The better performance of drop augmentations for single-task city classification implies that this augmentation is not fitting for scene classification, which in the multi-task model acts as a regulariser over the city classification task, reducing overall performance. Despite this robust performance by all augmentations, the cyclic and stretch single-task CNNs are not performing as well as the multi-task models, but the cyclic+stretch accuracy increases by $+6\%$ to $77\%$ accuracy.

Overall, cyclic augmentations have the best performance compared to all other augmentation methods. Drop augmentations appear to reduce the accuracy for ASC models leading us to believe that this type of augmentation might not be appropriate to preserve some important information for audio scenes. It is possible that a more conservative number of frames to drop would perform better, as in current experiments the number of frames can be up to the length of the recording.

\begin{figure}[bt]
    \centering
    \includegraphics[width=\columnwidth]{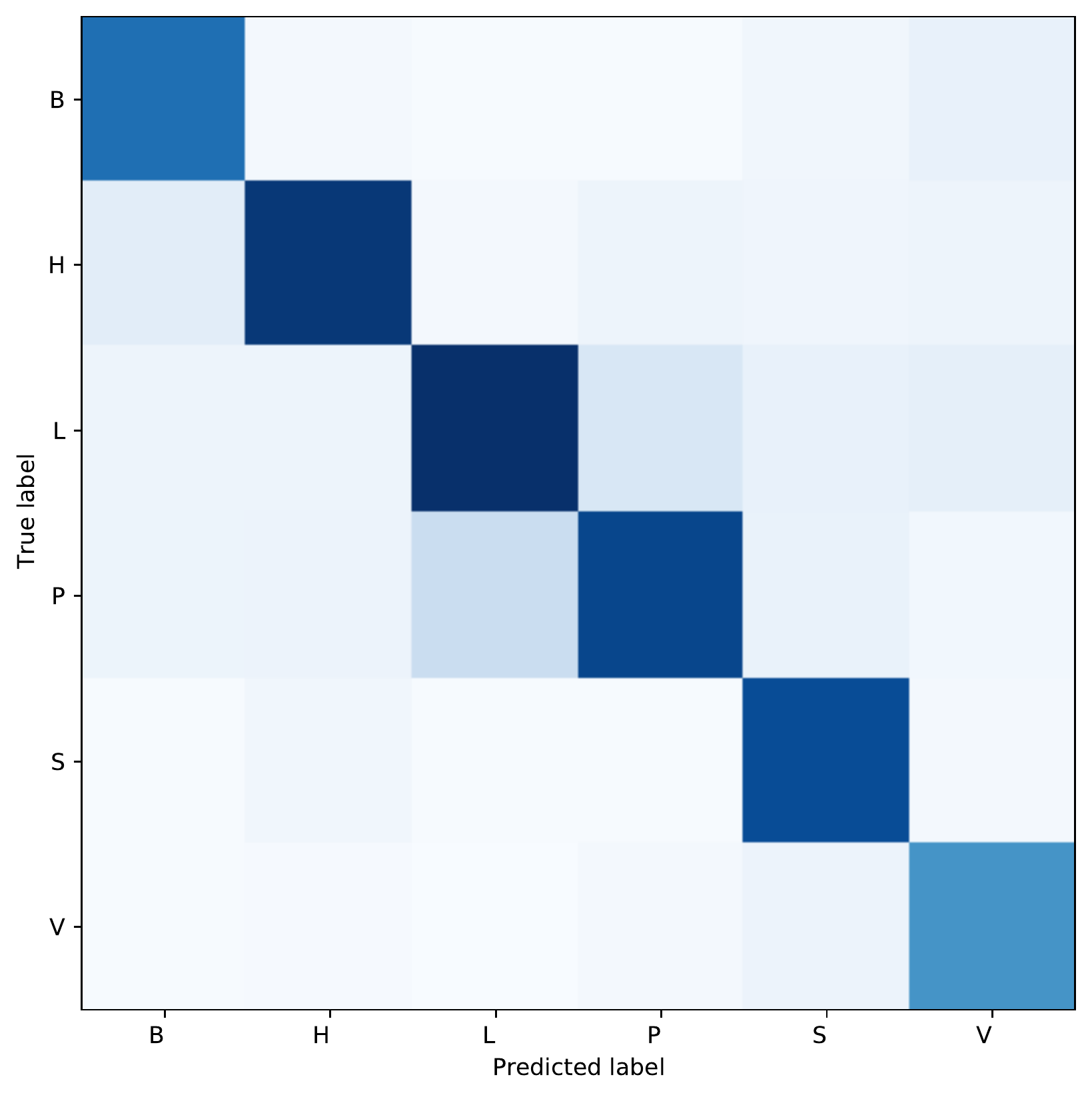}
    \caption{Confusion matrix for our best model (cyclic multi-task CNN). Classes are denoted by their first initial: B(arcelona), H(elsinki), L(ondon), P(aris), S(tockholm), V(ienna)}
    \label{fig:confus_mat}
\end{figure}

Our larger multi-task CNN architecture has overall a better performance for both scene and city classification compared to the CNN model proposed in \cite{bear2019city}. The best model for city classification is the larger multi-task model presented in Table~\ref{tab:bigparams} with an accuracy of $79\%$. Furthermore, except for the single-task model using all augmentations, the models using augmented data converged in less epochs than the baseline without any augmentation. Whilst this was not the original goal of this paper it is a useful observation for optimising training geotagging classifiers.

Figure~\ref{fig:confus_mat} shows a city-wise confusion matrix for the multi-task model trained using cyclic augmentations which scored $79\%$ accuracy in round two. The bulk of the accuracy comes from the London and Paris recordings which are very robustly classified. This is consistent with the observations in \cite{bear2019city} which uses the same dataset. Also consistent with these prior observations is that Vienna is the least well identified of the six cities.  

\begin{table}[!ht]
    \centering
    \caption{City classification test Accuracy (\%) by Scene with the best model}
    \begin{tabular}{l|r|r}
    \hline
    Scene & City Accuracy & \#testfiles\\
    \hline
         Airport         & $59$ & 265\\
         Bus             & $91$ & 242\\
         Metro           & $90$ & 261\\
         Metro Station   & $89$ & 259 \\
          Park            & $47$ & 242 \\
         Pedestrian Street & $74$ & 247\\
         Public Square   & $81$ & 216\\
         Shopping Mall   & $77$ & 279\\
         Traffic         & $86$ & 246\\
         Tram            & $91$ & 261\\
         
    \end{tabular}
    \label{tab:byScenePredictions}
\end{table}

Finally, it was observed in \cite{heise2020fuzz} that the difficulty of predicting cities can vary by the scene itself. Therefore, to understand the quality of our predictive models, we list the city classification accuracy of our best performing model for each scene in Table~\ref{tab:byScenePredictions}. The number of test files for each scene is in the last column. It is apparent that the \textit{airport} and \textit{park} scenes are having significant difficulty in separating out which cities they are recorded in with $59\%$ and $47\%$ accuracy, respectively. This might be due to the international/formulaic organisation and behaviour of airports all over the world, and the quiet outdoors nature of park recordings which are unlikely to contain distinctive features. This would be an area worthy of further research in the future for tackling audio geotagging in difficult scenes. 

\section{Conclusions and Future work}
\label{sec:conclusions}
In this work we have completed a comprehensive review of common data augmentation methods with a goal of improving audio geotagging (city classification on sound scenes). We have learned that applying the augmentations to stereo input instead of mono input and then mixing the two channels (with care taken for time alignment where necessary) is the optimal method of preparing the data for training. 


Our results suggest that while different augmentations work for each task in the multi-task model, for example \textit{drop} is not good for ASC but \textit{cyclic} is best for geotagging, using data augmentations overall can improve the performance for city classification. Also, supplementing training data with data augmentation enables faster converging of CNN geotagging classifiers. 

Other future work includes better model development and parameter optimisation, and testing on the latest twelve city dataset from DCASE 2019. As discussed in Section~\ref{sec:results} we will further evaluate augmentation parameters such as the \textit{drop} length. We will further improve scene specific geotagging for difficult scenes, such as \textit{park} and \textit{airport} with non-discriminative features.  

Ultimately, we have reviewed the most common audio data augmentation methods and shown that with \textit{cyclic} augmentation and a multi-task CNN, we have improved the state-of-the-art sound scene geotagging by $+23\%$ to $79\%$ accuracy.

\section{Acknowledgements}
VM is supported by the BBSRC grant  BB/R008736/1. EB is supported by a Turing Fellowship under the EPSRC grant EP/N510129/1.

\bibliographystyle{IEEEtran}
\bibliography{references}

\end{document}